\documentclass[preprint,authoryear,12pt]{elsarticle}

\usepackage{amsmath,amssymb}
\usepackage{bm}
\usepackage{algorithm}
\usepackage{algpseudocode}
\usepackage{booktabs}
\usepackage{hyperref}
\usepackage{graphicx}
\usepackage{geometry}
\newcommand{\R}{\mathbb{R}}
\newcommand{\bfa}{\bm{a}}
\newcommand{\bfc}{\bm{c}}
\newcommand{\bfPhi}{\bm{\Phi}}
\newcommand{\bfSigma}{\bm{\Sigma}}
\newcommand{\bfTheta}{\bm{\Theta}}
\newcommand{\bfxi}{\bm{\xi}}
\newcommand{\bfu}{\bm{u}}
\newcommand{\norm}[1]{\left\|#1\right\|}

\journal{arXiv preprint}

\begin{document}

\begin{frontmatter}

\title{Manifold-Adapted Sparse RBF-SINDy:\\
Unbiased Library Construction and Unsupervised Discovery of\\
Dynamical States in Turbulent Wall Flows}

\author[1]{M.~P\'{e}rez Cuadrado}
\author[1]{G.M.~Cavallazzi}
\author[1]{A.~Pinelli\corref{cor1}}
\cortext[cor1]{Corresponding author.
  \textit{Email:} Alfredo.Pinelli.1@citystgeorges.ac.uk}
\address[1]{School of Science and Technology, Department of Engineering,
  City St George's, University of London, London, UK}

\begin{abstract}
The turbulent attractor in wall-bounded flow is not a structureless strange set:
it has a skeleton of dynamically distinct states, connected by rare but directed
transitions, whose geometry is imprinted on the invariant measure of the
phase-space trajectory.
We show that this skeleton is recoverable from wall measurements alone ---
wall pressure and wall-shear stress --- without physical labels or prior
knowledge, provided the data-driven function library used to learn the dynamics
is constructed to respect the intrinsic geometry of the attractor rather than
the extrinsic variance hierarchy of the POD representation.
Standard sparse-identification approaches introduce two structural biases at the
library construction stage.
The steep energy decay of POD spectra causes the Euclidean distance in
$k$-means to be dominated by the leading modes, collapsing all basis-function
centres into a low-dimensional subspace and leaving the transition dynamics
without library coverage.
Separately, turbulent trajectories slow near quasi-invariant flow states, so
uniform time-stepping over-represents those states relative to the physical
invariant measure and under-samples the fast transitional passages between them.
Both biases are corrected by resampling the trajectory uniformly in arc-length
before clustering, and by replacing the Euclidean metric in $k$-means with the
Mahalanobis metric derived from the local covariance of each cluster.
The arc-length resampling drives the empirical measure towards the physical
invariant measure; the Mahalanobis metric produces ellipsoidal basis functions
whose support is adapted to the local Riemannian geometry of the attractor at
each cluster, rather than to the global variance hierarchy of the POD spectrum.
A single global sparse regression on this corrected library, with centre
allocation weighted by dynamical coherence and cluster population, completes
the model.
Applied to a minimal turbulent channel at $Re_\tau = 180$, the corrected
latent-space geometry renders the two phases of the near-wall autonomous cycle
directly visible to unsupervised clustering: stable streak attractors and
burst-initiating instabilities separate into two sharp populations without any
physical input, mapping onto the exact coherent structure skeleton of the flow.
This organisation is absent when standard biased approaches are used.
The learned model recovers the correct invariant measure, saturates the Lyapunov
predictability horizon set by the intrinsic chaos of the flow, and furnishes a
differentiable vector field on which invariant solutions can be located directly
by Newton iteration.
\end{abstract}


\end{frontmatter}

\section{Introduction}
\label{sec:intro}

The dynamics of near-wall turbulence are organised around a self-sustaining
cycle of elongated low-speed streaks, their sinuous instability, and a
subsequent breakdown and regeneration~\citep{jimenez1999autonomous}.
This cycle operates on timescales of order $100t^+$ and is accessible, at
least in part, through measurements at the wall: the pressure and shear-stress
footprint of the near-wall structures is coherent enough that reduced-order
models driven exclusively by wall quantities can capture a significant fraction
of the flow dynamics~\citep{jimenez1991minimal}.

Sparse identification of nonlinear dynamics
(SINDy)~\citep{brunton2016discovering}, combined with a POD encoding of the
wall-measurement snapshots, offers a natural framework for this problem.
One seeks a sparse ordinary differential equation
$\dot{\bfa} = \bfTheta(\bfa)\bfxi$ in the space of POD latent coefficients,
assembled from a library of candidate functions $\bfTheta$.
For turbulent flows, global polynomial libraries are insufficiently flexible,
and radial basis functions (RBFs) placed on the
attractor~\citep{reinbold2020using,loiseau2018constrained} provide a better
approximation basis.
The idea of partitioning the phase-space trajectory by $k$-means clustering
and building a reduced model on that partition was introduced by
\citet{kaiser2014cluster} as cluster-based reduced-order modelling (CROM),
and extended to continuous-time network dynamics by \citet{fernex2021cluster}.
Those approaches model inter-cluster dynamics by a discrete Markov transition
matrix; the present work replaces that discrete jump model with a continuous,
differentiable sparse ODE whose library is built from the same cluster
geometry, enabling trajectory integration and, as discussed below, direct
access to invariant solutions of the learned system.
A related but architecturally distinct line of work replaces the Markov model
with a collection of local intrusive ROMs, one per cluster, switching between
them via a nearest-centroid affiliation function.
\citet{colanera2025qlrom} demonstrate this quantized local ROM (ql-ROM)
strategy on the Kuramoto--Sivashinsky and Kolmogorov flow equations, showing
gains in numerical stability and long-time statistical accuracy over a single
global Galerkin model.
Because ql-ROMs require the governing equations for each local Galerkin
projection and introduce coordinate discontinuities at cluster boundaries ---
which the authors identify as a limitation requiring future treatment ---
they are not directly applicable to observation-limited settings such as the
wall-measurement problem considered here.
The present framework is non-intrusive throughout and produces a globally
smooth vector field, so no switching discontinuities arise and the full
benefits of a differentiable ODE are retained.

The difficulty is that the POD latent space of a turbulent flow is neither
flat nor isotropic.
POD spectra decay rapidly: the variance in the leading mode can exceed that
in the last retained mode by three or four orders of magnitude.
A $k$-means algorithm that uses the Euclidean distance on this space will
distribute its centres to minimise inertia in the high-variance directions,
leaving the low-variance modes --- which carry the inter-state transition
structure that governs the long-time dynamics --- without any RBF support.
A separate but equally damaging effect comes from the non-uniform speed of
turbulent trajectories.
Near quasi-invariant states such as the laminar streak phase, trajectories
slow substantially; uniform time-stepping therefore over-represents those
states relative to the physical invariant measure and under-samples the fast
bursting transitions.
When $k$-means is applied to uniformly sampled data, the cluster geometry
inherits this distortion, placing additional centres in the already
over-represented slow regions.

The consequences are not merely a degradation in regression accuracy.
Because both biases misrepresent the attractor geometry, the learned model
converges to the wrong invariant measure: it may reproduce short-time
trajectories reasonably well but fails to recover the correct long-time
energy statistics and the dynamical state structure of the flow.
The goal of the present work is to correct both biases at the level of library
construction, before any regression is performed, so that the resulting model
is consistent with the physical invariant measure from the outset.

The correction proceeds in two steps.
The trajectory is first resampled uniformly in arc-length, which redistributes
the sampling density in proportion to the local phase-space speed and thereby
approximates uniform sampling with respect to the invariant measure.
The $k$-means clustering is then performed with the Mahalanobis distance
derived from the local covariance of each cluster rather than the Euclidean
distance.
This stretches the effective distance along low-variance directions so that
cluster boundaries, and the RBF centres derived from them, are distributed
uniformly across the full latent-space geometry rather than compressed into
the subspace of the leading modes.
The resulting library feeds a single global sparse regression, with no local
models and no model switching.

When this corrected framework is applied to a minimal turbulent channel at
$Re_\tau = 180$, the clustering step reveals something not visible with
standard approaches: the trajectory organises into two sharply separated
populations in the space of cluster residence time and phase-space velocity
coherence, reproducing the streak and burst phases of the near-wall cycle
without any physical input.
These two populations are interpretable as the data-driven counterpart of the
exact coherent structure (ECS) skeleton that organises the turbulent
attractor~\citep{nagata1990three,cvitanovic2010geometry,kawahara2012significance}: the high-residence cluster
neighbourhood approximates the lower-branch laminar-streak ECS, while the
low-residence neighbourhood approximates the upper-branch saddle-type
structures that initiate the burst.

\section{Problem formulation}
\label{sec:setup}

\subsection{POD encoding}

\begin{figure}[t]
  \centering
  \includegraphics[width=0.85\textwidth]{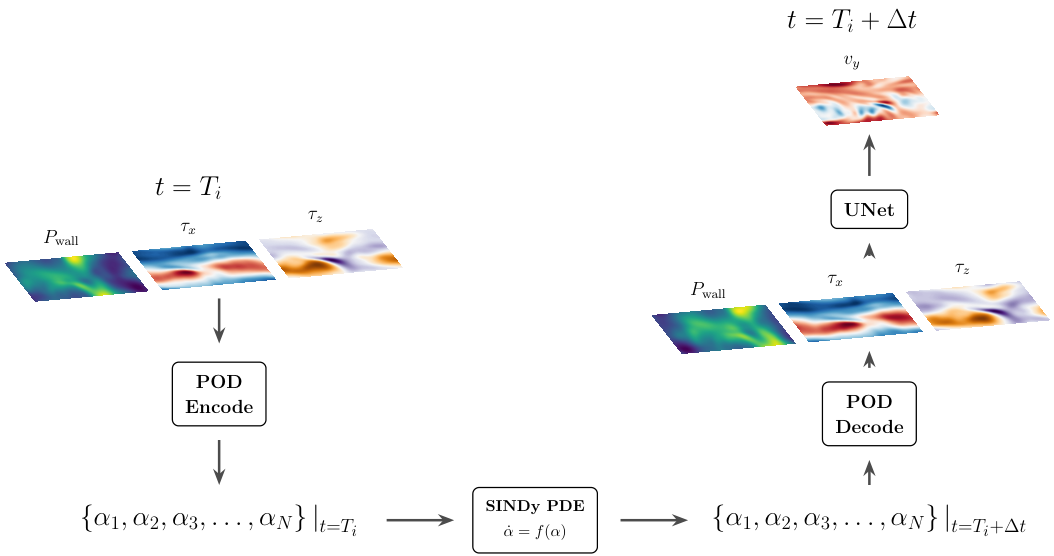}
  \caption{Schematic of the modelling pipeline. Wall measurements
    $\{P_\mathrm{wall},\tau_x,\tau_z\}$ at time $T_i$ are encoded onto
    POD latent coefficients $\{\alpha_j\}$, advanced by the learned
    RBF-SINDy ODE, and decoded back to wall fields at time $T_i + \Delta t$.}
  \label{fig:framework}
\end{figure}

The overall modelling pipeline is sketched in figure~\ref{fig:framework}.
Let $\bfu(\bm{x},t) \in \R^N$ denote the discretised wall-measurement field,
consisting of $P_\mathrm{wall}$, $\tau_x$, and $\tau_z$ sampled on both walls.
Given $M$ snapshots $\{\bfu_i\}_{i=1}^M$, the fluctuation matrix
$\bm{U}' = \bm{U} - \bar{\bfu}\bm{1}_M^\top$ is formed and a truncated
randomised SVD $\bm{U}' \approx \bfPhi_r \bm{\Sigma}_r \bm{V}_r^\top$ is
computed, retaining $r$ modes that account for a fraction $\eta \geq 0.99$
of the total variance \citep{sirovich1987turbulence}.
The latent state $\bfa(t) = \bfPhi_r^\top \bfu'(\cdot,t) \in \R^r$ encodes
the snapshot, and $\bfu' \approx \bfPhi_r\bfa$ decodes it.
Time derivatives $\dot{\bfa}$ are estimated by second-order centred finite
differences.

\subsection{Sparse identification in the latent space}

The dynamics of the latent state are approximated by
\begin{equation}
  \dot{\bfa} = \sum_{k=1}^{K} \xi_k\,\phi_k(\bfa),
  \label{eq:sindy}
\end{equation}
where $\{\phi_k\}_{k=1}^K$ are the $K$ library functions and
$\bfxi \in \R^{K \times r}$ is a sparse matrix of amplitude coefficients.
The library functions are fixed entirely by the data geometry, as described
in section~\ref{sec:correction}; SINDy operates only on the amplitudes,
solving the linear regression $\dot{A} \approx \bfTheta(A)\bfxi$
and enforcing sparsity by either sequentially thresholded least squares
(STLS), or penalising the regression with an $L_1$ term~\citep{brunton2016discovering}.
The library $\bfTheta$ consists of $K$ radial basis functions (RBFs)
centred at points $\{\bfc_k\}_{k=1}^K$ in the latent space,
augmented with the $r$ linear terms $\{a_j\}_{j=1}^r$.

Radial basis functions of the form $\phi_k(\bfa) = \varphi(\|\bfa -
\bfc_k\|)$ are a classical tool for approximating smooth functions on
$\R^r$~\citep{buhmann2003radial,wendland2004scattered}.
For any choice of strictly positive definite kernel $\varphi$, the set
$\{\phi_k\}_{k=1}^K$ spans a reproducing kernel Hilbert space (RKHS) and
can approximate any continuous function on a compact domain to arbitrary
accuracy as $K \to \infty$ with centres dense in the
domain~\citep{micchelli1986interpolation}.
In practice $K$ is finite and the centres are placed on or near the
attractor, so the library is adapted to the support of the invariant
measure rather than to a uniform grid.
The linear terms are retained to capture any residual affine structure and
to improve conditioning of the regression problem.

Among admissible kernels, the Gaussian RBF
\[
\varphi(d) = \exp\!\left(-\frac{d^2}{2\sigma^2}\right)
\]
is a natural choice in a dynamical-systems context. Here
$d = \|\bfa - \bfc_k\|$ denotes the Euclidean distance between the
state $\bfa$ and the centre $\bfc_k$ of the $k$-th kernel, and
$\sigma$ is the bandwidth controlling the spatial extent of the
basis function. This expression corresponds to the isotropic form
of the Gaussian kernel, in which the same width $\sigma$ is used
in all directions of the $r$-dimensional state space. 

The Gaussian RBF possesses several desirable properties:
{\em i)} it is infinitely differentiable, ensuring that the learned right-hand side 
$f(\bfa) = \bfTheta(\bfa)\bfxi$ is smooth and well posed for numerical integration;
{\em ii)} its Fourier transform is also Gaussian, providing spectral locality: each RBF term
contributes primarily within a well-defined neighbourhood of its centre,
without introducing spurious long-range interactions between distant centres;
{\em iii)} it aligns naturally with the local geometry of the data distribution. The advantages
of this feature will become particularly evident once the Mahalanobis metric is introduced in Section~\ref{sec:correction}.

Each Gaussian RBF is characterised by two sets of parameters: its centre
$\bfc_k \in \R^r$ and its width (or bandwidth) $\sigma_k \in \R^+$.
The centre determines where the basis function acts in state space,
while the width controls the spatial extent of its influence.

In the simplest isotropic case a single scalar width
$\sigma \in \R^+$ is shared across all $r$ directions. Each RBF then
requires $r+1$ parameters: the $r$ coordinates of the centre and one
bandwidth. In the anisotropic case the scalar width is replaced by a
full $r \times r$ symmetric positive-definite matrix $\bfSigma$,
allowing different spreads and orientations in state space; this
introduces $r + r(r+1)/2$ parameters per function.

In the present framework the centres $\bfc_k$ are determined by
arc-length discretisation of the latent trajectory, with spacing
weighted by the local phase-space speed $\|\dot{\bfa}\|$, so that
dynamically active regions receive a denser coverage of centres.
The width matrices are obtained from the per-cluster sample covariance
matrices $\bfSigma_k$, computed from the snapshots assigned to each
cluster and inverted to yield the precision matrices
$\bfSigma_k^{-1}$ that define the anisotropic kernel shape.

Because both centres and widths are fixed directly from the data
geometry before the regression step, the identification problem
in~(\ref{eq:sindy}) remains purely linear in the amplitude
coefficients $\bfxi$. This provides a significant computational and
algorithmic advantage over approaches that optimise kernel centres,
widths, and coefficients jointly, where the resulting problem is
nonlinear and considerably harder to condition.

The procedure used to construct these centres and precision matrices
is described in the next section.

\section{Methodology}
\label{sec:correction}

\subsection{Coordinate bias from Euclidean clustering}

The POD singular values satisfy $\sigma_1 \geq \sigma_2 \geq \cdots \geq
\sigma_r$ and $\mathrm{Var}(a_j) \propto \sigma_j^2$.
In turbulent channel flow the ratio $\sigma_1^2/\sigma_r^2$ typically exceeds
$10^3$.
The standard $k$-means algorithm minimises the total Euclidean inertia
\begin{equation}
  \mathcal{I} =
  \sum_{k=1}^K \sum_{i \in \mathcal{C}_k}
  \norm{\bfa(t_i) - \bfc_k}^2
  = \sum_{j=1}^r \sum_{k,i} (a_j(t_i) - c_{k,j})^2,
  \label{eq:inertia}
\end{equation}
and since each term in the $j$-sum scales with $\sigma_j^2$, the optimiser
invests all $K$ centres to resolve the first two or three directions.
The RBFs placed there have negligible support in the remaining directions,
which carry the transition dynamics between flow states and determine the
invariant measure.

\subsection{Temporal sampling bias}

The empirical measure from uniform time-stepping,
$\mu_\Delta = M^{-1}\sum_i \delta_{\bfa(t_i)}$, is related to the physical
invariant measure $\mu_\infty$ by
\begin{equation}
  \frac{d\mu_\Delta}{d\mu_\infty}(\bfa)
  \propto \norm{\dot{\bfa}(\bfa)}^{-1}.
  \label{eq:measure_bias}
\end{equation}
Regions where the trajectory is slow accumulate many snapshots; fast
transitional regions are sparsely sampled.
When $k$-means is applied to $\mu_\Delta$, centres are pulled towards the
over-represented slow regions and the regression targets are dominated by
them, systematically misrepresenting the dynamics that govern transitions
between attractors.

The two biases are independent and must each be corrected separately.
Coordinate-correcting methods such as ISOMAP change the extrinsic
representation of the manifold but leave the temporal sampling distribution
unchanged; arc-length resampling corrects the measure but not the metric.
Only their combination yields a library that is simultaneously unbiased in
geometry and in sampling density.


\subsection{Arc-length resampling}

The trajectory $\{\bfa(t_i)\}_{i=1}^M$ is resampled at $M$ equally spaced
points along its arc-length
\begin{equation}
  \ell_i = \sum_{j=1}^{i-1}\norm{\bfa(t_{j+1}) - \bfa(t_j)},
  \quad i = 1,\ldots,M,
  \label{eq:arclength}
\end{equation}
by interpolating at the uniform grid $\ell^*_n = n\,\ell_M/(M-1)$.
The resampled empirical measure $\hat{\mu}$ approximates the arc-length
measure $d\ell/\ell_M$ and satisfies
\begin{equation}
  \frac{d\hat{\mu}}{d\mu_\Delta}(\bfa) \propto \norm{\dot{\bfa}(\bfa)}^2,
  \label{eq:corrected_measure}
\end{equation}
up-weighting fast segments relative to slow ones.
For ergodic systems with bounded phase-space speed,
$\hat\mu \to \mu_\infty$ as $M \to \infty$, and all subsequent steps
operate on the resampled trajectory $\{\hat{\bfa}_n\}$.

\subsection{Arc-length discretisation and cluster assignment}

The arc-length resampled trajectory $\{\hat{\bfa}_n\}$ defines a
one-dimensional curve in $\R^r$.
This curve is discretised into $K$ segments by placing $K$ centre points
$\{\bfc_k\}_{k=1}^K$ at equally spaced arc-length positions along it,
with the spacing $\Delta s$ stretched locally in proportion to the
phase-space speed $\|\dot{\bfa}\|$ so that dynamically active regions
receive denser coverage.
Each resampled snapshot $\hat{\bfa}_n$ is then assigned to the nearest
centre in the Euclidean sense, defining $K$ disjoint clusters
$\mathcal{C}_k = \{n : k = \arg\min_j \|\hat{\bfa}_n - \bfc_j\|\}$.

For each cluster $k$ the sample covariance is computed from the assigned
snapshots,
\begin{equation}
  \bfSigma_k =
  \frac{1}{|\mathcal{C}_k|}
  \sum_{n \in \mathcal{C}_k}
  (\hat{\bfa}_n - \bfc_k)(\hat{\bfa}_n - \bfc_k)^\top + \varepsilon I_r,
  \label{eq:cov_k}
\end{equation}
where the regularisation $\varepsilon I_r$ prevents singularity in
under-populated clusters.
The precision matrix $\bfSigma_k^{-1}$ is obtained by direct inversion
and encodes the local geometry of the attractor at each centre: it
stretches distances along low-variance directions so that the support of
the corresponding RBF reflects the actual shape of the data cloud rather
than the global POD variance hierarchy.
The cluster count $K$ is determined automatically by
G-means~\citep{hamerly2003learning}, which tests each candidate cluster
for Gaussianity and subdivides it if the test fails.

\begin{figure}[t]
  \centering
  \includegraphics[width=0.85\textwidth]{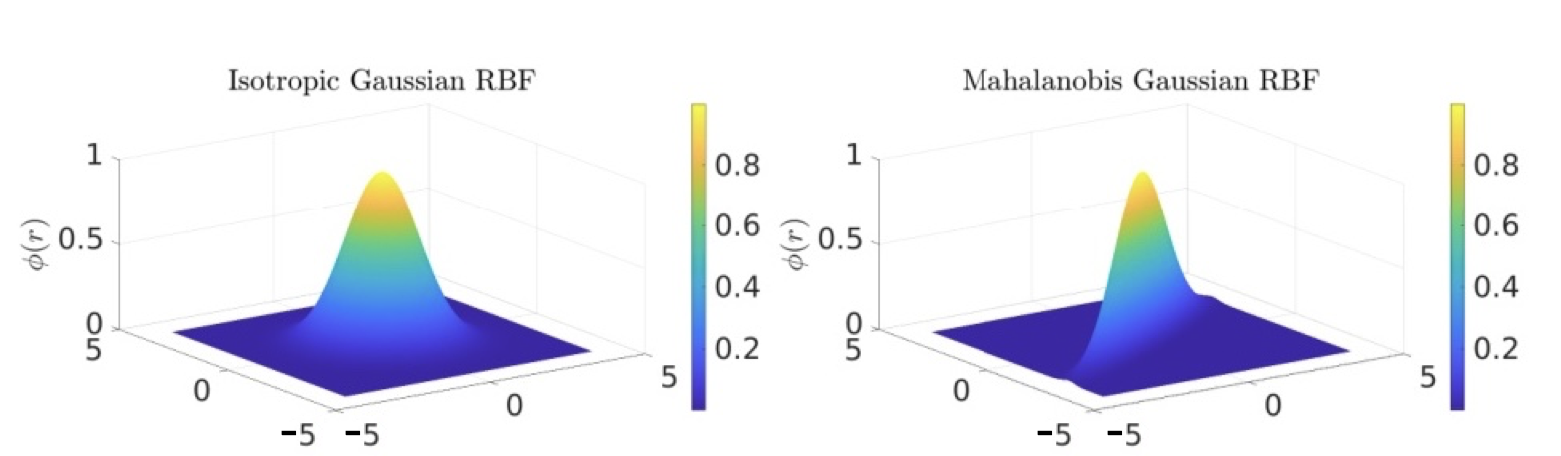}
  \caption{Comparison of isotropic Gaussian RBF (left) and Mahalanobis
    Gaussian RBF (right) for the same cluster centroid. The isotropic
    function assigns equal bandwidth in all directions regardless of
    the local data geometry; the Mahalanobis function stretches along
    low-variance directions and compresses along high-variance ones,
    matching the ellipsoidal shape of the cluster.}
  \label{fig:rbf_comparison}
\end{figure}

Figure~\ref{fig:rbf_comparison} illustrates the difference between isotropic
and Mahalanobis RBFs for representative clusters: the isotropic RBF
ignores the anisotropy of the cluster, while the Mahalanobis RBF aligns
its support with the local data distribution.

\begin{figure}[t]
  \centering
  \includegraphics[width=0.75\textwidth]{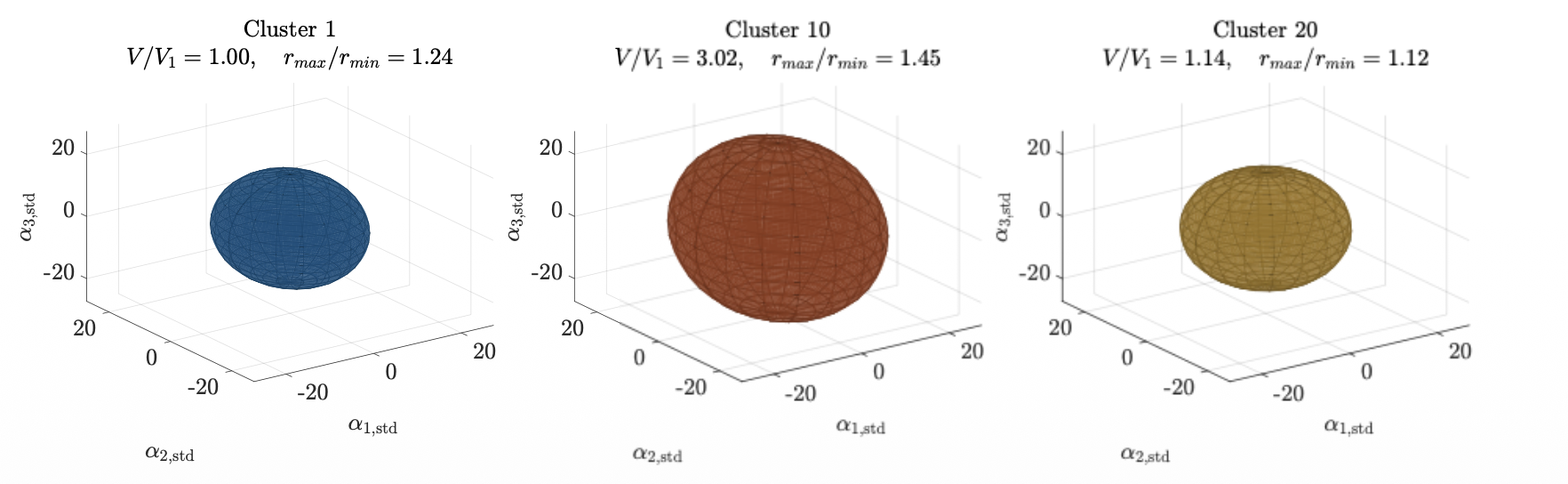}
  \caption{Three representative clusters (1, 10, 20) visualised in
    the space of the first three standardised POD coefficients.
    The ellipsoids represent the one-sigma Mahalanobis surface
    $(\bfa - \bfc_k)^\top\bfSigma_k^{-1}(\bfa-\bfc_k) = 1$.
    The varying size and orientation reflect the strongly anisotropic
    and heterogeneous geometry of the attractor.}
  \label{fig:cluster_ellipsoids}
\end{figure}

The $k$-th library function is the Mahalanobis Gaussian RBF:
\begin{equation}
  \phi_k(\bfa) =
  \exp\!\left(-(\bfa - \bfc_k)^\top \bfSigma_k^{-1}
    (\bfa - \bfc_k)\right).
  \label{eq:mahal_rbf}
\end{equation}
The level sets of $\phi_k$ are ellipsoids whose axes are aligned with
the eigenvectors of $\bfSigma_k$ and scaled by the corresponding
eigenvalues, so the support of each basis function is geometrically
congruent with the local data distribution it represents.
Figure~\ref{fig:cluster_ellipsoids} shows the one-sigma Mahalanobis
surfaces for three representative clusters, illustrating the strongly
anisotropic and heterogeneous geometry of the attractor that isotropic
RBFs would fail to capture.

\subsection{Coherence-weighted centre allocation}

The allocation of the centre budget $K$ across clusters follows
\begin{equation}
  K_k \propto (1 - c_k)\sqrt{N_k},
  \label{eq:allocation}
\end{equation}
where $N_k = |\mathcal{C}_k|$ and
\begin{equation}
  c_k =
  \frac{1}{|\mathcal{C}_k|}
  \sum_{i \in \mathcal{C}_k}
  \frac{\dot{\bfa}(t_i)\cdot\dot{\bfa}(t_{i+1})}
  {\norm{\dot{\bfa}(t_i)}\norm{\dot{\bfa}(t_{i+1})}}
  \label{eq:coherence}
\end{equation}
is the mean cosine similarity of consecutive phase-space velocity vectors
within cluster $k$.
Clusters with low coherence (rapidly changing trajectory direction) receive
more centres; larger clusters receive proportionally more coverage.
Within each cluster the centres are distributed by farthest-point sampling
in the Mahalanobis metric.
\subsection{Algorithm recap}

Offline training proceeds as follows.
A randomised SVD of the fluctuation snapshot matrix yields $\bfPhi_r$ and
the latent trajectory $A$.
Time derivatives $\dot A$ are estimated by centred differences.
The trajectory is resampled uniformly in arc-length to obtain $\hat A$,
which is then discretised into $K$ clusters (with $K$ from G-means) by
arc-length centre placement and nearest-neighbour assignment.
For each cluster the sample covariance $\bfSigma_k$ is computed and
inverted to give the precision matrix $\bfSigma_k^{-1}$, fully defining
the RBF library $\bfTheta$ via~(\ref{eq:mahal_rbf}).
Centre allocation follows~(\ref{eq:allocation}).
With the library fixed, STLS~\citep{brunton2016discovering} solves the
linear regression $\dot A \approx \bfTheta(A)\bfxi$ for the sparse
amplitude matrix $\bfxi$.

Online prediction from an initial wall-measurement snapshot $\bfu_0$ requires
the encoding $\bfa_0 = \bfPhi_r^\top(\bfu_0 - \bar{\bfu})$, followed by
RK4 integration of $\dot{\bfa} = \bfTheta(\bfa)\bfxi$, and decoding
$\bfu = \bfPhi_r\bfa + \bar{\bfu}$.
The cost per time step is $O(sr)$ for $s$ active RBF terms, compared to
$O(N\log N)$ for spectral DNS, giving online speedups of $10^3$--$10^5$.

\section{Results}
\label{sec:results}

\subsection{Dataset and POD encoding}

\begin{figure}[t]
  \centering
  \includegraphics[width=\textwidth]{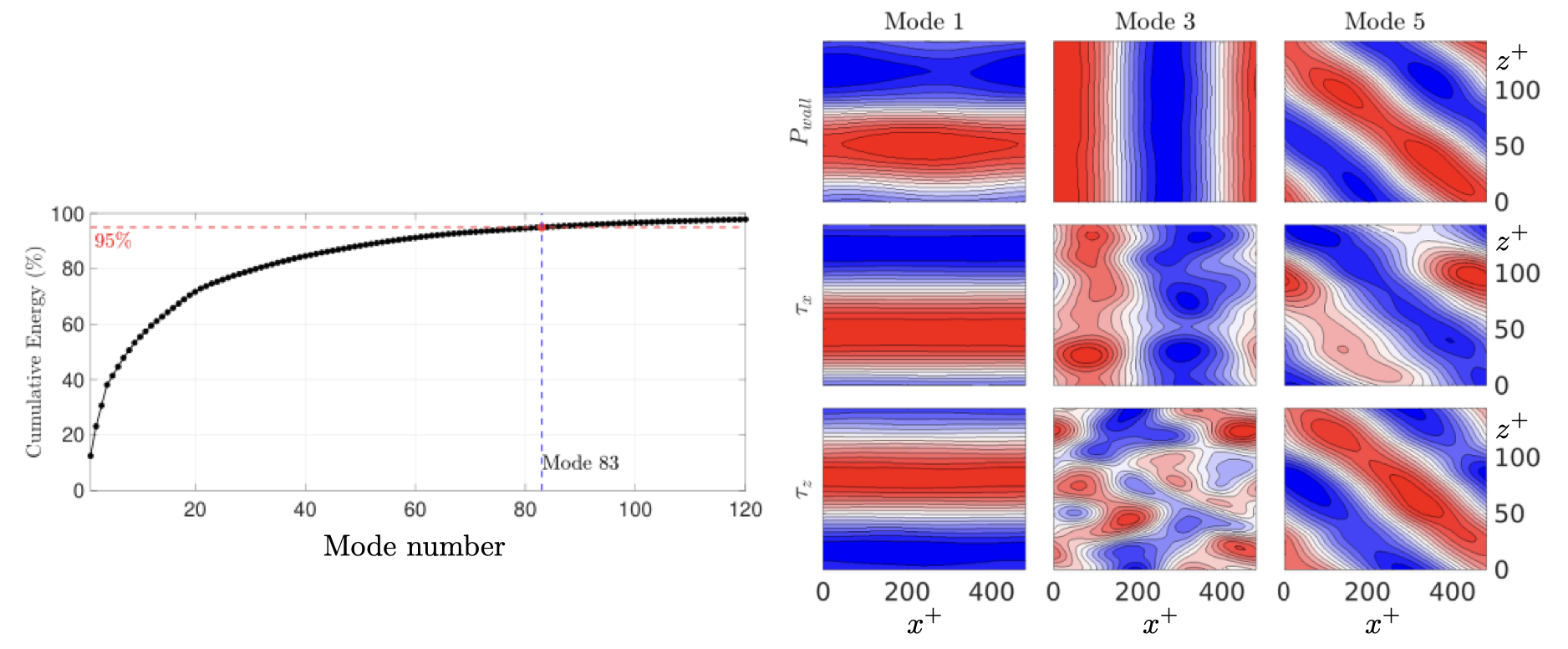}
  \caption{Left: cumulative energy as a function of POD mode index,
    with the 95\% threshold at mode 83 indicated.
    Right: wall-pressure $P_\mathrm{wall}$, streamwise shear $\tau_x$,
    and spanwise shear $\tau_z$ contours for modes 1, 3 and 5
    (columns), illustrating the physical content of the leading modes.}
  \label{fig:data_encoding}
\end{figure}

The training data consist of $M = 20{,}000$ snapshots of
$\{P_\mathrm{wall},\,\tau_x,\,\tau_z\}$ from a minimal turbulent channel at
$Re_\tau = 180$~\citep{jimenez1991minimal}, sampled at intervals of $0.5t^+$.
A randomised SVD retains $r = 83$ modes accounting for 95\% of the total
wall-measurement variance, as shown in figure~\ref{fig:data_encoding}.
Mode~1 reflects the large-scale pressure signature of elongated streaks;
Modes~3 and~5 capture the streamwise-periodic and oblique patterns
associated with streak instability.

\subsection{Dynamical states recovered by unsupervised clustering}

\begin{figure}[t]
  \centering
  \includegraphics[width=0.65\textwidth]{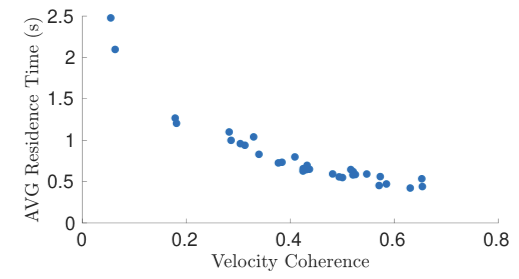}
  \caption{Average residence time versus velocity coherence for the
    32 clusters returned by G-means on the arc-length resampled,
    Mahalanobis-corrected trajectory.
    Two populations are clearly separated: high-residence / low-coherence
    clusters (upper left) correspond to quasi-steady streak states;
    low-residence / high-coherence clusters (lower right) correspond
    to burst-initiating instabilities.}
  \label{fig:residence_coherence}
\end{figure}

G-means clustering of the corrected latent trajectory returns $K = 32$
clusters.
For each cluster we compute the mean residence time and velocity
coherence~(\ref{eq:coherence}); the result is shown in
figure~\ref{fig:residence_coherence}.

\begin{figure}[h!]
  \centering
  \includegraphics[width=\textwidth]{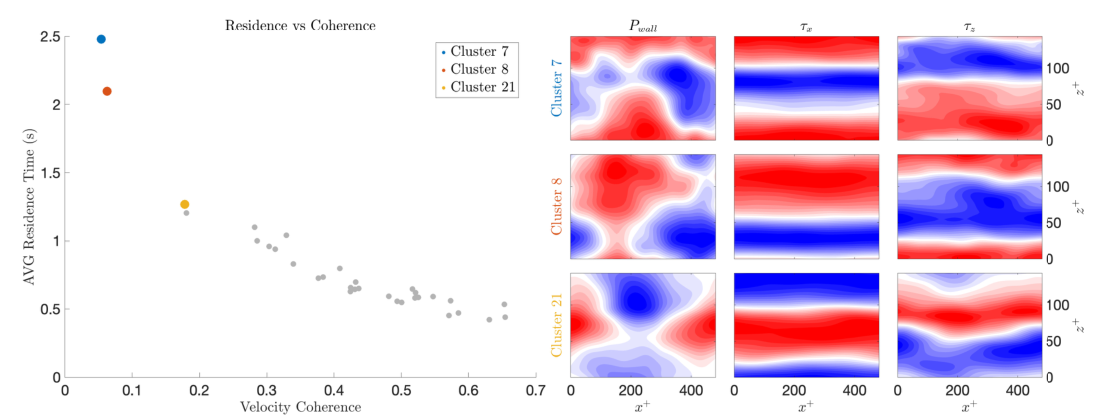}
  \caption{Three representative high-residence / low-coherence clusters
    (7, 8, 21). Left: their positions in the (coherence, residence time)
    plane. Right: reconstructed wall fields
    $P_\mathrm{wall}$, $\tau_x$, $\tau_z$ at each cluster centroid,
    showing the large-scale, spanwise-uniform streak signature.}
  \label{fig:stable_clusters}
\end{figure}

The distribution is sharply bimodal.
The first population, exemplified by clusters 7, 8 and 21
(figure~\ref{fig:stable_clusters}), has high residence time and low
velocity coherence.
The trajectory settles here for extended periods, with the phase-space
velocity changing direction erratically rather than following a coherent
path.
The wall fields reconstructed from the cluster centroids show large-scale,
nearly spanwise-uniform pressure distributions and strongly streamwise-aligned
wall-shear stress --- the signature of the quasi-steady elongated streak state
that constitutes the laminar phase of the near-wall
cycle~\citep{jimenez1999autonomous}.

\begin{figure}[t]
  \centering
  \includegraphics[width=\textwidth]{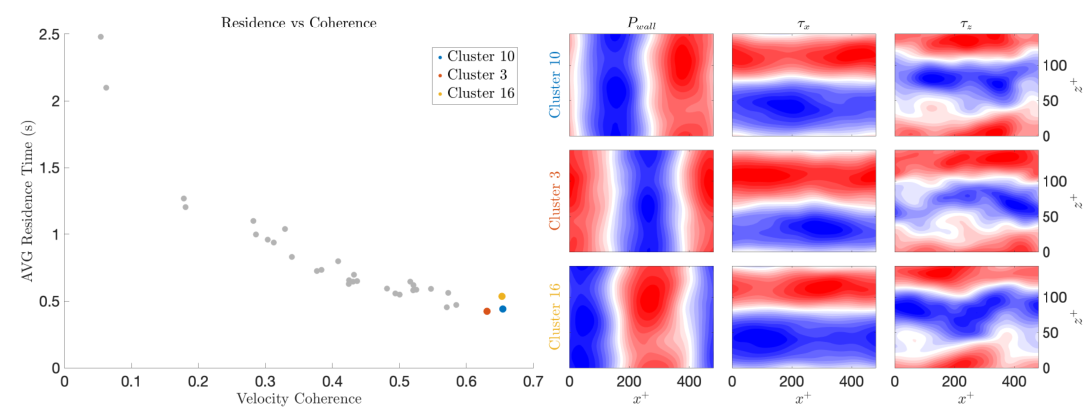}
  \caption{Three representative low-residence / high-coherence clusters
    (3, 10, 16). Left: their positions in the (coherence, residence time)
    plane. Right: reconstructed wall fields at each cluster centroid,
    showing the oblique, three-dimensional instability signature.}
  \label{fig:unstable_clusters}
\end{figure}

The second population, exemplified by clusters 3, 10 and 16
(figure~\ref{fig:unstable_clusters}), has short residence time and high
velocity coherence.
The trajectory passes through these states rapidly, with the phase-space
velocity strongly aligned from one snapshot to the next.
The wall fields are oblique and strongly three-dimensional, consistent
with the sinuous instability of streaks and the initiation of a turbulent
burst.

This bimodal organisation maps directly onto the streak and burst phases
identified by~\citet{jimenez1999autonomous} through a completely independent
analysis.
It emerges here with no physical assumptions, as a geometric consequence
of the corrected latent-space discretisation.
When the same clustering is applied to uniformly sampled data with the
Euclidean metric, the bimodal structure is absent: coordinate bias compresses
the centre distribution into the subspace of the leading modes and temporal
sampling bias inflates the apparent extent of the slow streak states,
obscuring the separation between the two populations.

\begin{figure}[h!]
  \centering
  \includegraphics[width=0.60\textwidth]{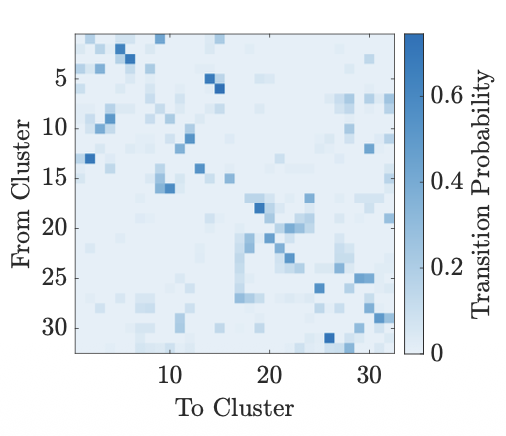}
  \caption{Cluster-to-cluster transition probability matrix $P_{kj}$,
    computed from the arc-length resampled trajectory.
    The block structure reflects the separation between
    stable streak clusters and unstable burst-initiating clusters,
    with directed transitions between the two populations.}
  \label{fig:transition_matrix}
\end{figure}

The two populations also admit a precise dynamical interpretation within the
exact coherent structure (ECS) framework for turbulent shear
flows~\citep{kawahara2012significance}.
That framework identifies the turbulent attractor with a skeleton of invariant
solutions --- equilibria, travelling waves, and unstable periodic orbits ---
connected by heteroclinic and homoclinic orbits, with the statistical
properties of turbulence recoverable as weighted averages over these solutions.
The high-residence / low-coherence clusters recovered here correspond to the
neighbourhoods of lower-branch ECS: laminar-like, low-dissipation states near
which the trajectory lingers for extended periods, analogous to the
streak-dominated lower-branch equilibrium of the near-wall
cycle~\citep{jimenez1999autonomous}.
The low-residence / high-coherence clusters correspond to saddle-type or
upper-branch ECS through which the trajectory passes rapidly and directedly on
its way to triggering a burst and returning to the streak attractor.
The precision matrices $\bfSigma_k^{-1}$ of the Mahalanobis clusters are
natural geometric descriptors of these neighbourhoods: the ellipsoidal
one-sigma surface of a high-residence cluster is elongated along the
directions of slow approach to and departure from the nearby invariant
solution, directly encoding the local stable and unstable manifold structure.
The cluster-to-cluster transition matrix (figure~\ref{fig:transition_matrix})
then provides the probabilistic coarse-graining of the heteroclinic connections
between these ECS families, constituting a directed Markov skeleton of the
near-wall cycle entirely consistent with what the ECS framework predicts on
theoretical grounds.
That this structure emerges from wall-observable data alone, without access to
the interior velocity field, reflects the wall-attached character of the
relevant ECS at $Re_\tau = 180$: the pressure and shear-stress footprint of
the near-wall structures is coherent enough to make their phase-space
neighbourhoods distinguishable in the POD latent space of the wall
measurements.

The cluster-to-cluster transition probability matrix $P_{kj}$
(figure~\ref{fig:transition_matrix}) exhibits a clear block structure, with
rare but directed transitions from streak-state clusters to burst-initiating
clusters and rapid return.
This directed Markov skeleton of the autonomous cycle is a natural output of
the corrected clustering.
The coherence-weighted allocation concentrates centres in the
burst-initiating clusters, which have low coherence and high dynamical
complexity, consistently with the weighting prescribed by
equation~(\ref{eq:allocation}).

\begin{figure}[h!]
  \centering
  \includegraphics[width=0.8\textwidth]{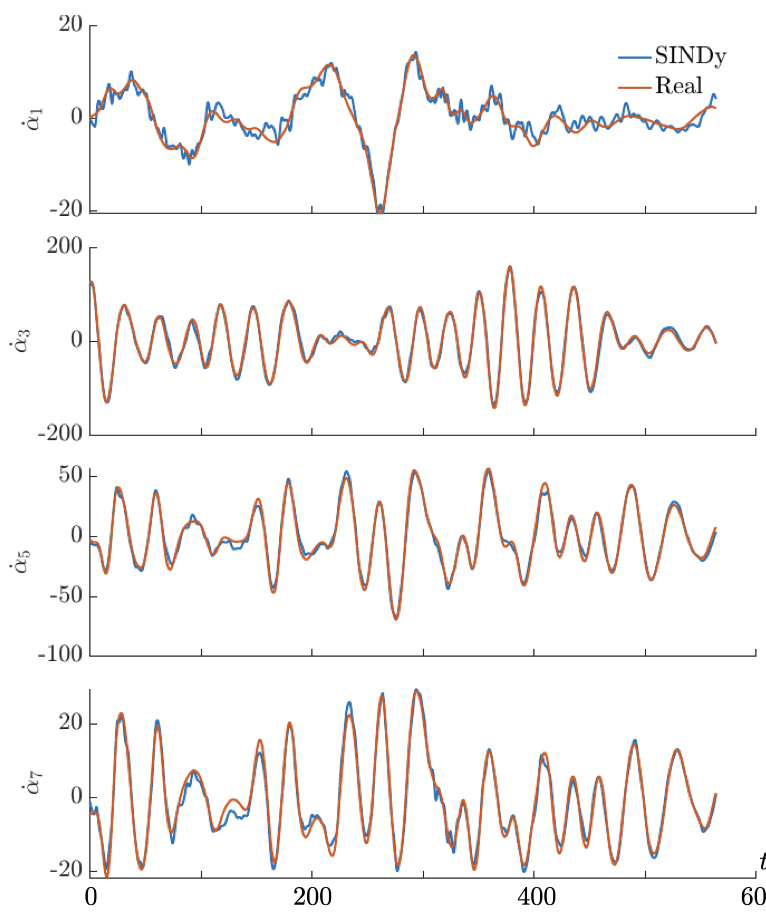}
  \caption{Time derivatives of latent coefficients $\dot\alpha_j$
    for modes $j = 1, 3, 5, 7$ (rows): ROM integration (blue) versus
    DNS finite differences (orange) over $t^+ \in [0, 600]$.
    The ROM reproduces the dominant oscillations and burst excursions
    across all retained modes without low-frequency drift or amplitude
    inflation.}
  \label{fig:mode_comparison}
\end{figure}

\subsection{Mode time-traces: ROM versus DNS}

The most direct test of the learned ODE is a comparison of the temporal
evolution of individual latent coefficients $\alpha_j(t)$ against the
DNS reference over a contiguous integration window.
Figure~\ref{fig:mode_comparison} shows the time derivatives
$\dot\alpha_j$ produced by the ROM alongside the finite-difference
values from the DNS for modes $j = 1, 3, 5, 7$ over $t^+ \in [0, 600]$,
corresponding to approximately six near-wall cycle periods.
The agreement is close across all four modes throughout the window,
with the ROM capturing the amplitude and phase of the dominant
oscillations as well as the sharper excursions associated with burst
events visible as isolated large-amplitude dips in $\dot\alpha_1$ near
$t^+ \approx 50$ and $t^+ \approx 320$.
The higher modes $\dot\alpha_5$ and $\dot\alpha_7$ show somewhat
larger pointwise errors at burst events, consistent with the
expectation that energy cascades to higher POD modes during the
burst phase and the Lyapunov divergence is fastest there.
Crucially, the ROM does not exhibit the low-frequency drift or
amplitude inflation that characterises poorly conditioned RBF
regressions, confirming that the Mahalanobis library construction
and the STLS sparsification together produce a well-conditioned model.

\begin{figure}
  \centering
  \includegraphics[width=0.75\textwidth]{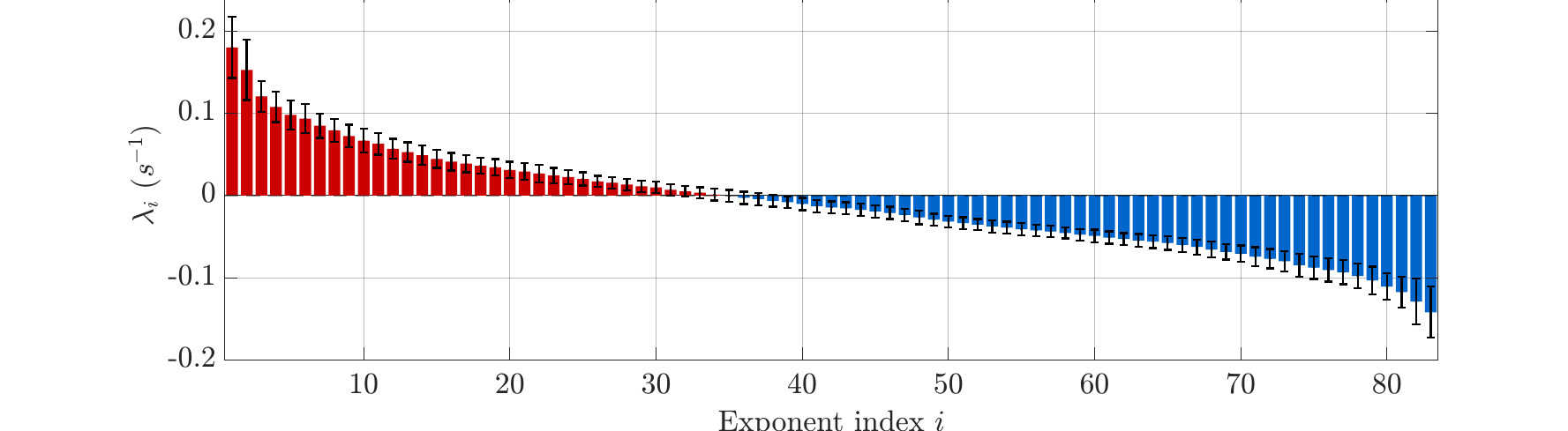}
  \caption{Lyapunov spectrum of the learned ROM, ordered by exponent
    index. Positive exponents (red) define the chaotic core; negative
    exponents (blue) define the stable complement. Error bars indicate
    variability across trajectory segments. The leading exponent
    $\lambda_1 \approx 0.2\,\mathrm{s}^{-1}$ sets the predictability
    horizon $t^* \approx 5t^+$.}
  \label{fig:lyapunov}
\end{figure}

\subsection{Dynamical fidelity of the learned model}

The central question for any data-driven ROM is whether it has learned
the correct dynamics or merely fitted the training trajectory.
Three complementary diagnostics address this.

The first is invariant measure recovery.
When the learned ODE is integrated forward from arbitrary initial
conditions, the marginal PDFs of the latent coefficients, the turbulent
kinetic energy spectrum, and the mean and variance of the wall-shear
stress all match the training statistics, and the cluster-to-cluster
transition probabilities of the integrated trajectory agree with those
computed directly from the DNS data.
The model has converged to the correct invariant measure, not to a
trajectory that happens to pass through the training snapshots.

The second diagnostic is the Lyapunov spectrum.
Figure~\ref{fig:lyapunov} shows the full spectrum of the ROM.
Approximately 35 modes carry positive exponents, constituting the
chaotic core of the dynamics; the remainder are stable, confirming
that the model reproduces sparse chaos consistent with the near-wall
turbulence and justifying the sparse SINDy representation.
The leading exponent $\lambda_1 \approx 0.2\,\mathrm{s}^{-1}$ sets the
theoretical predictability horizon $t^* \sim 1/\lambda_1 \approx 5t^+$,
beyond which any deterministic prediction must diverge from the true
trajectory regardless of model quality.

\begin{figure}
  \centering
  \includegraphics[width=\textwidth]{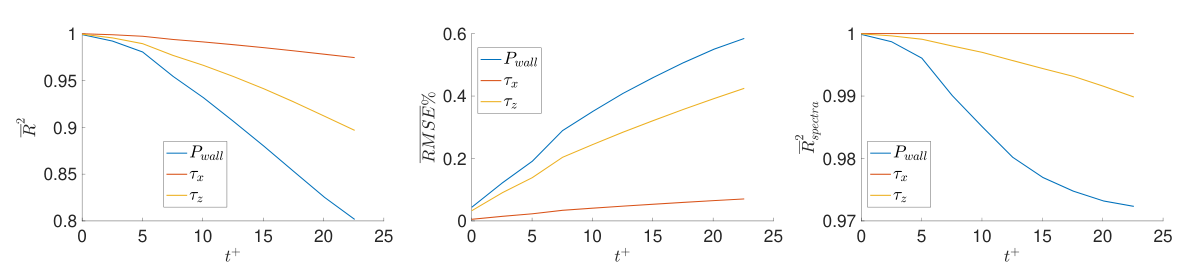}
  \caption{Prediction accuracy as a function of time horizon.
    Left: Pearson correlation $R^2$ for $P_\mathrm{wall}$ (blue),
    $\tau_x$ (red), $\tau_z$ (yellow).
    Centre: normalised RMSE for the same three fields.
    Right: radial spectral correlation $R^2_\mathrm{spectra}$,
    showing that large-scale structures are preserved throughout
    the predictability window.
    The vertical dashed line marks the Lyapunov horizon $t^*$.}
  \label{fig:prediction_accuracy}
\end{figure}

The third diagnostic is short-time prediction accuracy interpreted
against the Lyapunov horizon.
Figure~\ref{fig:prediction_accuracy} shows that wall-field predictions
achieve Pearson correlation $R^2 > 0.97$ for all three quantities at
horizons up to $t^*$, with spectral correlation above 0.98 across all
resolved wavenumbers at $10t^+$.
Beyond $t^*$ the prediction error grows and saturates at a level
consistent with the Lyapunov spectrum, confirming that the error growth
is governed by the intrinsic chaos of the flow rather than by any
deficiency of the model.
Taken together, the three diagnostics establish that the corrected
RBF-SINDy framework has recovered not just a fitting of the training
data but a faithful reduced-order representation of the governing
dynamics.

\section{Conclusions}
\label{sec:conclusions}

Building SINDy libraries by isotropic $k$-means clustering of uniformly
sampled POD trajectories introduces two structural biases that corrupt the
learned invariant measure, and that have not previously been identified or
corrected in the sparse-identification literature.
The first arises from the steep energy decay of POD spectra, which causes
Euclidean $k$-means to concentrate all basis-function centres in the
subspace of the leading modes, leaving the transition dynamics without
library coverage.
The second arises from the non-uniform speed of turbulent trajectories,
which causes uniform time-stepping to over-represent quasi-invariant slow
regions relative to the physical invariant measure.
Correcting both requires treating the latent space as a Riemannian manifold:
arc-length resampling adjusts the sampling measure so that the empirical
distribution approximates the physical invariant measure, and
Mahalanobis-metric clustering adjusts the distance metric to reflect the
local geometry of the data at each point on the attractor rather than the
global variance hierarchy of the POD spectrum.
The two corrections are independent and both necessary; their combination
yields a library that is geometrically unbiased, without introducing local
model switching or any discontinuity in the learned vector field.

The consequences are demonstrated on a minimal turbulent channel at
$Re_\tau = 180$, driven entirely by wall measurements --- wall pressure and
wall-shear stress --- with no access to the interior velocity field.
The most striking result is qualitative: the corrected latent-space geometry
renders the dynamical state structure of near-wall turbulence directly
visible to unsupervised clustering, without any physical labelling or prior
knowledge.
The 32 clusters returned by G-means organise into two sharply separated
populations in the plane of velocity coherence and mean residence time.
The high-residence, low-coherence population corresponds to the
quasi-steady elongated streak states of the laminar phase of the
autonomous cycle; the low-residence, high-coherence population corresponds
to the burst-initiating instability states through which the trajectory
passes rapidly and directedly before returning to the streak attractor.
This bimodal organisation maps precisely onto the picture established by
\citet{jimenez1999autonomous} from independent analysis, and is entirely
absent when the same clustering is applied to standard biased data.
The directed Markov skeleton encoded in the cluster-to-cluster transition
matrix provides, for the first time from wall-only data, a probabilistic
coarse-graining of the heteroclinic connections between these dynamical
state families.

The learned sparse ODE inherits the geometric fidelity of the corrected
library.
The model recovers the correct invariant measure of the flow, with long-time
statistics --- marginal PDFs of the latent coefficients, turbulent kinetic
energy spectrum, wall-shear variance, and transition probabilities ---
matching those of the DNS training data.
Short-time prediction accuracy, measured by Pearson correlation and spectral
correlation of the wall fields, remains high within the predictability
horizon $t^* \approx 5t^+$ set by the leading Lyapunov exponent of the ROM;
beyond $t^*$ the error growth rate is consistent with the Lyapunov spectrum,
confirming that the error is governed by the intrinsic chaos of the flow
rather than any deficiency of the model.
The framework thus saturates the theoretical predictability limit extractable
from the wall signal, a property that biased libraries fail to achieve
because they converge to the wrong attractor geometry.

The approach is not specific to wall turbulence.
Any dynamical system whose attractor is anisotropic in the representation
coordinates, or whose natural invariant measure is poorly approximated by
uniform time-stepping, will benefit from the same two corrections.
The combination of arc-length resampling and Mahalanobis-metric library
construction provides a principled, computationally lightweight route to
geometry-aware sparse identification in those settings, with no additional
hyperparameters beyond those already required by standard SINDy.

A natural continuation of this work is to close the connection with the
exact coherent structure (ECS) framework more precisely.
Because the RBF-SINDy model produces an explicit, differentiable right-hand
side $f(\bfa)$, fixed points of the latent-space ODE can be located by
Newton iteration on $f(\bfa^*) = 0$ at cost $O(sr)$ per evaluation, and
unstable periodic orbits can be sought by shooting methods or Poincar\'{e}
section analysis on the $r$-dimensional phase space --- computations that are
feasible in the ROM but wholly intractable on the full Navier--Stokes system.
The Jacobian $\partial f/\partial\bfa\!\mid_{\bfa^*}$ gives direct access to
the stable and unstable manifold dimensions at each fixed point and enables
a quantitative comparison with the ECS solutions documented by
\citet{kawahara2012significance}.
By lifting ROM fixed points to approximate full-field states via
$\bfu^* = \bar{\bfu} + \bfPhi_r\bfa^*$ and verifying their Navier--Stokes
residual through a data-driven wall-to-field
decoder~\citep{perez2026latent}, it becomes possible to assess how closely
the Mahalanobis cluster centroids proxy true invariant solutions of the
governing equations, constituting a route to ECS extraction from
observation-limited, wall-only data with full-field verification carried
out entirely within the data-driven pipeline.

\bibliographystyle{elsarticle-harv}
\bibliography{references}

\end{document}